\documentclass{aa}
\usepackage{graphicx}
\usepackage{natbib}

\def\lapp{\mathbin{\raise2pt \hbox{$<$} \hskip-9pt \lower4pt \hbox{$\sim$}}}
\def\gapp{\mathbin{\raise2pt \hbox{$>$} \hskip-9pt \lower4pt \hbox{$\sim$}}}

\newcommand{\ee}{\'{e}}
\newcommand{\kmps}{\;{\rm km\;s^{-1}}}
\newcommand{\cmmt}{\;{\rm cm^{-3}}}
\newcommand{\gcmmt}{\;{\rm g\;cm^{-3}}}

\begin{document}

\title{Nonradial and nonpolytropic  astrophysical outflows}
\subtitle{IX. Modeling T Tauri jets with a low mass-accretion rate}

\titlerunning{Nonradial and nonpolytropic  astrophysical outflows IX.}

 \author{C. Sauty
          \inst{1}
  \and
          Z. Meliani
          \inst{1,2}
  \and    J.J.G. Lima
          \inst{3,4}
  \and    K. Tsinganos
          \inst{5}
  \and    V. Cayatte
          \inst{1}
 \and    N. Globus
          \inst{1}
         }

  \offprints{{\tt christophe.sauty@obspm.fr}}

  \institute
    {Laboratoire Univers et Th\'eories, Observatoire de Paris, UMR 8102 du CNRS, Universit\'e Paris Diderot,
    F-92190 Meudon,  France
    \and
    Centrum voor Plasma Astrofysica, Celestijnenlaan 200B bus 2400, 3001 Leuven, Belgium
   \and
   Centro de Astrof\'{\i}sica, Universidade do Porto, Rua das Estrelas, 4150-762 Porto, Portugal
   \and
   Departamento de F\'{\i}sica e Astronomia, Faculdade de Ci\^{e}ncias,
        Universidade Porto, Rua do Campo Alegre, 687, 4169-007 Porto, Portugal
   \and IASA and Section of Astrophysics, Astronomy \& Mechanics,
        Department of Physics, University of Athens,
        Panepistimiopolis GR-157 84, Zografos, Greece
        }
  \date{Received 14/01/2011 / accepted 03/05/2011}

 \abstract
{A large sample of T Tauri stars exhibits optical jets, approximately half of which rotate slowly,   only at ten per
cent of their breakup velocity. The disk-locking mechanism has been shown to be inefficient to explain this
observational fact.}
{We show that low mass accreting T Tauri stars may have a strong stellar jet component that
can effectively brake the star to the observed rotation speed.}
{By means of a nonlinear separation of the variables in the full set of the MHD equations we construct
semi-analytical solutions describing the dynamics and topology of the stellar component of the jet that emerges
from the corona of the star.}
{We analyze two typical solutions with the same mass loss rate but different magnetic lever arms and jet
radii. The first solution with a long lever arm and a wide jet radius effectively brakes the star and can be applied to
the visible jets of T Tauri stars, such as RY Tau. The second solution with a shorter lever arm and a very narrow
jet radius may explain why similar stars, either Weak line T Tauri Stars (WTTS) or Classical T Tauri Stars (CTTS)
do not all have visible jets.
For instance, RY Tau itself seems to have different phases that probably depend on the activity of the star. }
{First, stellar jets seem to be able to brake pre-main sequence stars with a low mass accreting rate. Second, jets may be visible
only part time owing to changes in their boundary conditions.  We also suggest a possible scenario for explaining the
dichotomy between CTTS and WTTS, which rotate faster and do not have visible jets.
}


\maketitle

\section{Introduction}

\subsection{The angular momentum problem, observational facts}

Observations of star-forming regions show that several T Tauri stars are associated with well collimated
jets. These jets are made of strongly accelerated plasma that tra\-vels at a few
hundred km/s \citep{Bally09} and is magnetically collimated at large distances \citep{Dougadosetal00}.
The jets are usually associated with Classical T Tauri Stars (CTTS), which are low mass stars $\le 2 M_{\odot}$
in their late stages of pre-main sequence evolution, showing strong evidence of the presence of a surrounding
accretion disk. The outflow is usually thought to be ejected from the Keplerian disk
(e.g. \citeauthor{CabritAndre91}, \citeyear{CabritAndre91}).
However, a significant part of the jet, if not all, may be ejected by the star itself, at least for the lower mass
accreting T Tauri stars.

Observations also reveal that approximately half of the T Tauri stars rotate slowly, at about $10\%$ or less of their
breakup speed \citep{Mattetal10}. This indicates that a very efficient mechanism is at work to remove
the angular momentum in these stars; the nature of this mechanism is still controversial.

\cite{Kundurthyetal06},  see also \cite{Edwardsetal93},  claimed that  CTTS are slow rotators with circumstellar disks, while
WTTS are fast rotators without accretion disks.
However, the association of CTTS with slow rotators and WTTS with fast ones is not as yet observationally confirmed.
Moreover, \citeauthor{Stassunetal99} (\citeyear{Stassunetal99}, \citeyear{Stassunetal01}) and
\cite{Rebulletal04} could not find a strong correlation of fast rotators with stars that have dispersed
their disks. Conversely, new results  (\citeauthor{Rebulletal06}, \citeyear{Rebulletal06};
\citeauthor{Herbstetal07}, \citeyear{Herbstetal07}; \citeauthor{CiezaBaliber07}, \citeyear{CiezaBaliber07})
found a bimodal distribution where stars with disks rotate slower
than stars without disks. However, the two distributions clearly overlap.
The prevailing theory has been that magnetic interaction between the star and its circumstellar disk
regulates the stellar rotation periods. Attempts to observationally confirm  this conjecture have produced mixed results.
Hence, from the observational point of view,
the presence of an accretion disk may not be enough to explain stellar angular momentum removal.

In order to understand slow stellar rotators, \cite{Schatzman62} suggested that the magnetic braking of a stellar wind would be
sufficient even with a low mass loss rate, and this has been explored by several authors (e.g. 
\citeauthor{WeberDavis67}, \citeyear{WeberDavis67}, \citeauthor{Mestel68a}, \citeyear{Mestel68a}, \citeyear
{Mestel68b}). Following this track, it is essential to address stellar wind models in detail, with
and without the presence of an accretion disk, to see how efficient the wind is in extracting angular momentum from
the star. Moreover,  recent observations confirm  that stellar winds are present in at least $60\%$ of CTTS
\citep{Kwanetal07}, as was already suggested by various authors (e.g. \citeauthor{Edwardsetal03}, \citeyear{Edwardsetal03}).

 \cite{Decampli81} showed that  {\it thermal} stellar winds cannot support jets with a high mass loss rate,
simply because this would require too high temperatures to be physical.
For low mass loss rates, less  than $10^{-9}M_\odot/yr$, he showed
that a thermally driven or line-driven radiative wind can support the formation of the jet. For a higher mass loss rate,
such a driving is still possible provided that there is an extra supply of pressure from Alfv\'en waves.
If the temperature of the wind is around one million degrees,
the pressure can drive the jet up to 300  km/s. However, the pressure may exceed
the thermal pressure by a large amount  if it  includes ram pressure or
turbulent magnetic pressure.  The situation is similar to the well known case of the solar wind, where 
to obtain wind speeds of several hundred km/s with a temperature of a few million degrees
one requires extra pressure from either turbulence or kinetic effects.
These effects yield an effective  temperature of around ten million degrees (see \citeauthor{Aibeoetal07},
\citeyear{Aibeoetal07} and references therein).

Ignoring the important question regarding the nature of the pressure and following the early study of \cite{Decampli81}, most authors
disregarded stellar wind models up to now. This explains why disk wind models have been extensively
studied. 

\subsection{Disk wind theory}

In disk wind models, the main source of acceleration is the magneto-centrifugal driving, as
proposed originally by \cite{BlandfordPayne82}, who presented the first MHD
radial self-similar solutions extending the earlier hydrodynamical
solutions of \citeauthor{BardeenBerger78} (\citeyear{BardeenBerger78}) for galactic winds. Since then,
their original model has been improved in various ways
(e.g., \citeauthor{Lietal92}, \citeyear{Lietal92}; \citeauthor{Contopoulos94}, \citeyear{Contopoulos94} ;
\citeauthor{VlahakisTsinganos98}, \citeyear{VlahakisTsinganos98}; \citeauthor{Vlahakisetal00}, \citeyear
{Vlahakisetal00}).
In particular \citeauthor{Ferreira97} (\citeyear{Ferreira97}) has shown that a consistent treatment of the
connection with the disk is possible.  From the theoretical point of view however, self-similar disk wind models
are not consistently describing the inner part of the jet close to its axis \citep{Graciaetal06}.
The solution has to be cut at the inner and outer edges of the disk. From the observational point of view,
cold disk winds turned out to be too rapid and too light (\citeauthor{Garciaetal01b}, \citeyear
{Garciaetal01b}). This last point can be overcome with warm disk solutions
and explains the high speed, dense powerful jets of early TTauri stars like DG Tau, as well as their rotation
(\citeauthor{Ferreiraetal06}, \citeyear{Ferreiraetal06}). DG Tau is probably one of the most
powerful sources with an accretion rate of $10^{-6}M_\odot/$yr and a corresponding mass loss rate that is
about 10\% of this value \citep{Hartiganetal95}. In this case, disk wind solutions seem more adapted.
Recent observations suggest that DG Tau is variable on short timescales with an accretion rate dropping down to
$10^{-7}M_\odot/$yr \citep{Becketal10}.

Simultaneously, several numerical simulations have investigated the jet launching of these disk winds (e.g.,
\citeauthor{OuyedPudritz97}, \citeyear{OuyedPudritz97}; \citeauthor{Ustyugovaetal00},
\citeyear{Ustyugovaetal00}; \citeauthor{Krasnopolskyetal03}, \citeyear{Krasnopolskyetal03};
\citeauthor{CasseKeppens04}, \citeyear{CasseKeppens04}). By investigating various boundary conditions and initial
configurations, they confirmed that disk winds could indeed be accelerated
and collimated by their own magnetic field.
More recently several authors (\citeauthor{Matsakosetal08}, \citeyear{Matsakosetal08};
\citeauthor{Stuteetal08}, \citeyear{Stuteetal08}; \citeauthor{Matsakosetal09}, \citeyear{Matsakosetal09})
have shown that analytical solutions such as the radially self-similar ones are
structurally stable. This result indicates that these solutions are good complementary tools to numerical simulations.

According to theoretical arguments \citep{Pudritz&Norman86} and numerical simulations (e.g.  \citeauthor
{Melianietal06}, \citeyear{Melianietal06}), disk winds can remove most of the angular momentum of the
accreting plasma. However, disk winds need the presence of an inner stellar jet component to be globally consistent.

From the theoretical point of view,  it was initially proposed that fieldlines of the stellar magnetosphere anchored into the disk
could efficiently brake the central star, in the so-called disk-locking mechanism proposed by \cite{ChoiHerbst96}.
In principle a fieldline rooted into the star will slow down (respectively speed up) the central star rotation, if it connects
to a region of the disk rotating at lower rotational speed (respectively higher speed) after (respectively before) the
corotation radius. Using a simple analytical model based on the Ghosh \& Lamb mechanism (\citeyear{GhoshLamb79a},
\citeyear {GhoshLamb79b}), \citeauthor{MattPudritz05} (\citeyear{MattPudritz05})
showed however that the disk-locking efficiency is
reduced because it leads to an opening of the fieldlines that
disconnects the disk from the star. Instead they proposed that angular momentum can be lost
via accretion-powered stellar winds provided the mass loss rate is 1 to 10 \% of the accretion rate as usually inferred from
observations \citep{Calvet98}.
Their results were confirmed via numerical simulations (\citeauthor{MattPudritz08a}, \citeyear{MattPudritz08a},
\citeyear{MattPudritz08b}; \citeauthor{Mattetal10}, \citeyear{Mattetal10}).
\cite{Kuekeretal03}, using similar simulations, concluded that even when the disk-magnetosphere connection is not disrupted, the
magnetic field could slow down the central object but the dominant disk torque still spins up the star. As mentionned by
\cite{Romanovaetal09}, these conclusions are not definitive because long-term simulations are required to establish the validity of
this result. Ultimately the authors insist on the essential role of the axial jet.
\cite{ZanniFerreira09} performed various additional simulations where the disk-locking is not disrupted by the building of large
toroidal field.
However, even in this situation, the magnetic accretion torque is not sufficient to cause an efficient spin-down of the star.

\subsection{Stellar jet  theory}

As a conclusion of the precedent subsection, the presence of a pure stellar wind in addition to a possible outer disk wind is
required to effectively remove the excess of  stellar
angular momentum.

Besides, more evolved T Tauri stars like RY Tau seem to have weaker
jets with accretion rates on the order of only a few $10^{-8}M_\odot/$yr, which may be an indication that
disk winds are not always absolutely necessary for the weaker and more evolved jets.
We see that with a mass loss rate of only 10\% of this value, these
jets are more difficult to analyze. From spectroscopic observations of RY Tau, \citeauthor{GomezVerdugo01} (\citeyear
{GomezVerdugo01}, \citeyear{GomezVerdugo07}) suspected the presence of a small-scale pure stellar jet because the
UV emission lines originate in a region that is too small to be produced by a disk wind.
New observations (\citeauthor{StongeBastien08}, \citeyear{StongeBastien08}; \citeauthor{AgraAmboageetal09},
\citeyear{AgraAmboageetal09}) clearly show a microjet in this faint object.
It is interesting to
investigate if these (micro-)jets can be modeled through stellar winds or a
combination of a stellar wind and a sub Keplerian disk wind.

This combination of a two-component outflow is under numerical investigation
(\citeauthor{Mattetal03}, \citeyear{Mattetal03}; \citeauthor{Koide03}, \citeyear{Koide03}; \citeauthor{Kuekeretal03},
\citeyear{Kuekeretal03}; \citeauthor
{Matsakosetal08}, \citeyear{Matsakosetal08}). Recent simulations
(\citeauthor{Melianietal06}, \citeyear{Melianietal06}; \citeauthor{Romanovaetal09}, \citeyear{Romanovaetal09};
\citeauthor{Matsakosetal09}, \citeyear
{Matsakosetal09}; \citeauthor{Fendt09}, \citeyear{Fendt09})
also show the interplay of mixing between the two components.
Several authors, e.g. \cite{Kuekeretal03}, \cite{Romanovaetal09} and \cite{Fendt09}, have shown in their numerical simulations
that the disk-magnetosphere connection produces strong intermittent outflows, the results of flares and reconnection.
This is somewhat similar to coronal mass ejections in
the solar wind and in microquasars. Though important to model, those outflows are not necessarily more important on the long-term
duration compared to the continuous underlying steady flow, as suggested by \cite{Romanovaetal09}. Moreover, 
\cite{Matsakosetal09} have shown that the variability of the source does not destroy the continuous steady jet. Indeed, they
use the first solution presented in this paper, as initial condition for the inner spine jet of their simulations, but use a
polytropic equation of state. In their previous paper \citep{Matsakosetal08} they have shown that changing from non polytropic to
polytropic would reduce the size of the radius of the jet, without drastically affecting the outflow behavior, however.

However, numerical simulations are usually time-consuming to such an extent that they cannot explore a
wide range of parameters. Moreover, they do not really start ejecting mass directly
from the star but rather at a given height between the sonic and the Alfv\'{e}n surfaces. Therefore,
it is of absolute interest to model the stellar jet that originates in the central star.

All these problems can be overcome by studying MHD solutions obtained via a nonlinear separation of 
the variables. These so-called self-similar solutions contain as special cases all known MHD outflow solutions (e.g.,
\citeauthor{VlahakisTsinganos98}, \citeauthor{STT02}, \citeyear{STT02}, hereafter STT02).
As shown in \cite{ST94}, hereafter ST94, these self-similar models can be seen as a combination of
stellar wind and X-wind models. Conversely to formal X-winds (\citeauthor{Shuetal94}, \citeyear{Shuetal94};
\citeauthor{Shangetal02}, \citeyear{Shangetal02}), they do not require that all fieldlines emerge from a single
point of the disk in the form of a fan. Instead there is a smooth transition between the stellar wind and the
disk jet. They are also shown to
be structurally stable (\citeauthor{Matsakosetal08}, \citeyear{Matsakosetal08}; \citeyear{Matsakosetal09}). This was far from
obvious because the heating function is not polytropic as usually in most
analytical solutions. In a series of papers --
\cite{STT99} hereafter STT99, STT02, and \cite{STT04}, hereafter STT04 -- we have systematically explored the
full parameter space of this problem. In ST94 we have
shown that the double disk and star component was promising. However, the  solutions
presented there had mass loss rates that were too low. Here we present solutions where the stellar
jet is both under-dense and under-pressured compared to the disk
wind in the launching region. These solutions have the advantage to
adapt themselves not only to CTTS, which are connected with
the surrounding accretion disk, but also to WTTS
where there is { either no accretion disk or no connection with the disk itself} \citep{Walteretl88, Bertout89}.
Although winds from WTTS are difficult to measure, mass loss rates between $10^{-11}$ and  $10^{-10}M_\odot/$yr
have been obtained by  \cite{Andreetal92}. Winds of WTTS might be
similar to the solar wind in terms of collimation (see \citeauthor{Aibeoetal07}, \citeyear{Aibeoetal07}, and references therein).
However, these young stars have higher mass loss rates and rotational speeds than our  sun and might as well have strongly
collimated jets { independently of the presence of an accretion disk}.
Their rotation periods range from  0.6  to 24 days with peaks near two and eight days
\citep{Herbstetal07} while the solar rotational period is 24.5 day. Thus,  WTTS may well produce
self-collimated jets that nevertheless are so weak that they cannot be detected.

We will use the ST94 meridionally self-similar model to examine the contribution of the stellar
component to the overall jet and how efficiently it brakes the star. We first explore how  the model parameters can
be constrained from the observations of CTTS jets (Sect. \ref{sec.3}).
Then, we focus our attention on two different classes of solutions (Sections \ref{sec.4} and \ref{sec.5}) and explore
the efficiency of these winds to spin down the central object. In the
final section (Sect. \ref{sec.6}) we discuss how these two classes of solutions may explain the dichotomy
between CTTS and WTTS and how they can be adapted to also
model multiphases of T Tauri faint jets, such as in RY Tau.

\section{Meridional self-similar solutions}
\label{sec.2}

We summarize the main assumptions of our meridionally ($\theta-$)
self-similar treatment of the MHD equations in Appendix  \ref{A} . More details can be found in
STT94, STT99 and STT02. 

\subsection{Parameters and variables \label{sec21}}

Using spherical coordinates ($r, \theta, \varphi$), all quantities are normalized to the
Alfv\'en surface along the rotation axis, which is by assumption a sphere of radius $r=r_*$. 
The dimensionless radial
distance is denoted by $R=r/r_*$, while $B_*$, $V_*$ and $\rho_*$ are the
poloidal magnetic field, velocity and density along the polar axis at the
Alfv\'en radius $r_*$, with $V^2_*= B^2_* / 4 \pi \rho_*$.

The original system of MHD equations reduces to two coupled partial differential equations for the density and the magnetic flux.
Then, the momentum  conservation law provides three ordinary differential equations, which together with Eq.
(\ref{F}) can be solved for the four variables $M^2(R)$, $F(R)$, $\Pi(R)$ and
$G(R)$, which are  the square of the poloidal Alfv\'en Mach number, the expansion factor, the dimensionless pressure, and the dimensionless cross section radius, respectively.

The model is controlled by the following four parameters.
\begin{itemize}

\item
The  parameter $\delta$ governs the non spherically symmetric distribution of
the density. For $\delta > 0$ ($\delta < 0$) the density
increases (decreases) by moving away from the polar axis.

\item
The parameter $\kappa$ controls the non spherically symmetric
distribution of the pressure. For $\kappa > 0$ ($\kappa <0$) the gas pressure
increases (decreases) by moving away from the polar axis.

\item
The parameter $\lambda$ is related to the rotation of the  poloidal
streamlines at $R=1$ (i.e., at the Alfv\ee n surface).

\item
The parameter $\nu$ is defined by the gravitational field whose acceleration can be
written as
\begin{equation}
\vec{g} = - {{\cal G M}  \over r^2} \, \hat r
= -{1 \over 2} {V^2_* \over r_*}
{\nu^2 \over  R^2 }\, \hat r
\,,
\label{grv1}
\end{equation}
where ${\cal M}$ is the central gravitating mass. This
extra parameter $\nu$ is the ratio of the escape velocity to the poloidal flow
velocity on the polar axis at $R=1$ (i.e., at the Alfv\ee n surface),
\begin{equation}
\nu^2 ={2{\cal G M}   \over r_* V^2_*}
\,.
\label{grv2}
\end{equation}
\end{itemize}

\subsection{ Collimation efficiency of the magnetic rotator}

By integrating the momentum equations along a fieldline, we obtain the
conserved total energy flux density per unit of mass flux density. This is
equal to the sum of the kinetic and gravitational energies together with the
enthalpy and net heating along a specific streamline. In the framework of the
present meridionally self-similar model, the variation of the energy from one
line to the other gives an extra parameter (STT99):
\begin{eqnarray}
\epsilon  ={M^4\over (GR)^2}\left[ {F^2\over 4} - 1 \right]
- \kappa {M^4\over G^4}
- {(\delta\,- \kappa) \nu^2 \over R}
\nonumber\\
\label{EnEps}
+ {\lambda^2 \over G^2} \left({M^2-G^2 \over 1-M^2}\right)^2
+ 2\lambda^2{1-G^2 \over 1-M^2}\,\label{epsilon}
\,,
\end{eqnarray}
which is a constant on { all} streamlines (ST94).

Physically, $\epsilon$ is related to the { variation} across the
fieldlines of the specific energy that is left available to collimate the
outflow once the thermal content converted into kinetic energy and into
balancing gravity has been subtracted (STT99).

We can express $\epsilon / 2\lambda^2$  in terms of the conditions at the
source boundary $r_o$ (see STT99 for details),
\begin{equation}
{\epsilon  \over 2\lambda^2} =
{  E_{{\rm {Poynt.}},o} + E_{{\rm R},o}
+\Delta E_{\rm G}^*
\over E_{\rm {MR}}}
\,,
\end{equation}
where $E_{\rm {MR}}$ is the energy of the magnetic rotator (see Eq. 2.5a
in STT99), $E_{{\rm {Poynt.}},o}$
is the Poynting energy and $E_{{\rm R},o}$ is the rotational energy at the base.
$\Delta E_{\rm G}^*$ is the excess or deficit of the gravitational energy
(per unit mass) that is not compensated by the thermal driving, on a nonpolar
streamline compared to the polar one,
\begin{equation}
\Delta E_{\rm G}^*
= - {{\cal G}{\cal M} \over r_o}
\left[  1-{T_o(\alpha)\over T_o({\rm{pole}})} \right]
=  -{{\cal G}{\cal M} \over r_o} {(\delta - \kappa )\alpha
\over 1 + \delta \alpha }
\,.
\end{equation}

For $\epsilon >0$ collimation is mainly provided by magnetic terms, while for
$\epsilon <0$ the outflow can be confined only by thermal pressure, which is
not possible for overpressured flows. Accordingly, in STT99 we classified
flows with positive or negative $\epsilon$ as {\bf efficient} or {\bf inefficient} magnetic rotators,
respectively ({\bf EMR} or {\bf IMR}).

Considering the fairly large opening of stellar jets, we expect to find values of $\epsilon$
close to zero. We have shown in STT99 that this
corresponds to large asymptotic jet radius of magnetically confined outflows.

\section{Constraining the model parameters from the observations}
\label{sec.3}

\subsection{Stellar constraints}

Though their jets are fainter than those of class I and early CTTS, low mass accreting classical T Tauri stars
exhibit jets as many other class II objects.  They have well measured mass, radius and rotation frequency
\citep{Hartiganetal95, Bouvieretal97, HerbstMundt05, Marilli07}.
We suspected that one good such example of a faint jet in an evolved CTTS would be, among others, the RY
Tau jet (\citeauthor{GomezVerdugo01}, \citeyear{GomezVerdugo01}). This
has been recently observed and extensively studied (\citeauthor{StongeBastien08}, \citeyear{StongeBastien08};
\citeauthor{AgraAmboageetal09}, \citeyear{AgraAmboageetal09}).
Classical T Tauri stars are low mass stars between 0.5 and 2 solar masses, with typical radius
ranging from 2 to 3 solar radii (\citeauthor{Bertout89}, \citeyear{Bertout89}; \citeauthor{Moraetal01}, \citeyear{Moraetal01}).
They are typical precursors of solar-type stars. Rotational speeds are usually assumed
to be around one tenth of their breakup speed, namely typically around $15\kmps$ for CTTS and around $20\kmps$ for WTTS
(\citeauthor{Bouvieretal93}, \citeyear{Bouvieretal93}), but with large variations
and uncertainties. In particular, there is a large discrepancy between spectroscopic measurements of projected rotational
velocities and periods measured using photometry.

\subsection{Jet constraints}

For some CTTS we have measurements of various physical quantities in the optical jet itself
\citep{Hartiganetal95, Lavalley-Fouquetetal99, Cabrit07, StongeBastien08, AgraAmboageetal09}.
Observations give mass loss rates from $10^{-7}$ down to  $10^{-10}M_\odot/$yr.
On the other hand, low mass accreting stars are also associated with low mass loss rates, close to the minimum value.
The averaged asymptotic speeds are on the
order of $100\kmps$ to $300\kmps$  \citep{Bally09}. Microjets are hardly resolved transversely.
Thus, this speed might be a  mixing of the high- and low-velocity components similarly to what is observed in DG Tau  (\citeauthor
{Andersonetal03}, \citeyear{Andersonetal03}).
For the high-velocity component we can easily take a higher value and assume $400\kmps$ along the polar axis, where we
expect to find this maximum value. This remains of course arbitrary but is consistent with the model developed here.
In that respect, more precise measurements from
observations are also needed to assess the velocity profile in the jet (\citeauthor{Guntheretal09},
\citeyear{Guntheretal09}).

We assume that the jet carries a typical mass loss rate of a few times $10^{-9}\, M_\odot/$yr. Following
\cite{GomezVerdugo01}, we took  $3.1\times 10^{-9}\, M_\odot/$yr for the mass loss rate of RY Tau.
Electronic densities can be  directly measured only in shocks. Typically they are around $10^{4}\cmmt$. However, the inferred
electronic density is not unambiguously determined.

First the jet radius may be much smaller than the shock radius, as shown in
many numerical simulations (e.g. \citeauthor{Matsakosetal09}, \citeyear{Matsakosetal09}). As mentioned in the introduction,
\cite{Matsakosetal09} used as initial condition our solution for the inner spine jet but with different thermodynamics. Switching to a
polytropic equation of state reduces the radius of this inner part of the outflow by a significant factor, as shown in
\cite{Matsakosetal08}. Even with such a small radius of the spine jet, the long term variability of the stellar mass loss rate produces
knots ten times larger than the jet radius, typically around 5 to 10 AU. We expect that keeping the turbulent heating along the flow
would produce even larger knots.

Second, not all electrons are accelerated to sufficiently high energies.
Large uncertainties may also arise because we do not precisely know the ionization fraction of the jet.
Thus, densities and temperatures cannot be calculated precisely. The solutions presented in this paper can give a fairly 
accurate description of the outflow dynamics. In order to make a full synthetic emission map, further work is necessary starting from
the dynamics given from these solutions, calculating the amount of mechanical heating and self-consistently deducing the emission
lines by tuning the ionization fraction and the kinetic temperature. This has been done by \citeauthor{Garciaetal01a} (\citeyear
{Garciaetal01a}, \citeyear{Garciaetal01b}) for disk winds and will be similarly done for our solutions in a forthcoming study.

Finally, to have efficient magnetic braking, we looked for MHD solutions with long magnetic lever arms. We assume that a
typical lever arm, the Alfv\'en cylindrical radius, is around 10 times the stellar radius,
$\varpi_a=10\,\varpi_o$. For the meridionally self-similar model that we use this gives approximately
$r_\star=10\,r_o$ or $R_o=0.1$.

As shown in \cite{Spruit97}, if the accreting mass rate is $\dot M_{\rm acc}$, the angular momentum is
accreted at a rate
$\dot J_{\rm acc}=\dot M_{\rm acc}\varpi_o^2\Omega$. Similarly, the angular momentum loss rate in the wind is
$\dot J_{\rm wind}=\dot M_{\rm wind}\varpi_a^2\Omega$, because the plasma corotates up to the  Alfv\'en cylindrical radius
$\varpi_a$. If the entire
angular momentum from the disk is removed by the wind, we can equal these quantities $\dot J_{\rm wind}= \dot J_{\rm
acc}$. Combining the above we get
\begin{equation}
\varpi_a=\varpi_o\sqrt{\frac{\dot M_{\rm acc}}{\dot M_{\rm wind}}}
\,.
\end{equation}
If $\varpi_a=10\,\varpi_o$, we deduce from the above equation that the mass loss rate is
only 1\% of the mass accreting rate, which is a fairly low value. The wind is probably more efficient than this 1\% at extracting
mass from the disk. This means that even a relatively small wind outflow can remove angular momentum from the disk very
efficiently.

Alternatively, the wind can also remove angular momentum from the star if the magnetic fieldlines and streamlines are anchored
in the stellar photosphere instead of the disk. In that case, we can calculate the braking time to remove the stellar angular
momentum as
\begin{equation}
\tau= \frac{J_{\rm star}}{\dot J_{\rm wind}},
\end{equation}
where
\begin{equation}
{J_{\rm star}}=k \Omega M_{\rm star} r_{0}^2,
\label{Jstar}
\end{equation}
and
\begin{equation}
{\dot J_{\rm wind}}=\frac{\lambda r_{*} V_{*} {\dot M_{\rm wind}}}{2}\frac{\alpha_{\rm out}^2}{\psi_{\rm out}}
\label{Jdotwind}
\,,
\end{equation}
with $k$ the dimensionless inertial constant of the star \citep{Mestel68}. For Eq. (\ref{Jdotwind}) we used Eq. (3.19b) of
\cite{ST94}.

We use $k=0.06$ as for the Sun \citep{SchwartzSchubert69}, where 90\% of its mass is within $50\%$ of its radius
\citep{CoxAllen00}. For a T Tauri star with a bigger convective zone \citep{Bouvieretal97}, $k$ may be larger, up to
0.2. This maximum value is obtained for a fully convective star.

Substituting $\Omega$ from \cite{ST94} in Eq. (\ref{Jstar}), we obtain the following expression for
the braking time of the star
\begin{equation}
\tau=\frac{2 k M_{\rm star} r_{0}^2}{r_*^2 {\dot M_{\rm wind}}}
\frac{\psi_{\rm out}}{ {\alpha_{\rm out}}^2 \sqrt{1+\delta \alpha_{\rm out} }}
\label{tau}
\,,
\end{equation}
with
\begin{equation}
\psi_{\rm out} = \frac{2}{3\delta} [(1+\delta\alpha_{\rm out})^{3/2}-1]
\label{psiout}
\,.
\end{equation}

For instance, we know that CTTS have lifetimes on the order of a million years, while WTTS last on this phase around ten million
years.
The magnetic braking should accordingly be efficient in removing most of the accreted and stellar angular momentum during this
period, independent of the uncertainty factor we have on.

\subsection{Stellar parameters}
\label{sec.3.3}

To summarize, we will use a mass loss rate, stellar mass, radius and rotation given respectively by
\begin{itemize}
\item $\dot M_{\rm wind} = 3.1\times 10^{-9}M_\odot/$yr\,,
\item ${\cal M} = 1.5\,M_\odot$\,,
\item $r_o = 2.4\,r_\odot$\,,
\item $\Omega = 5.99\,\, 10^{-6}$ rad/s or $V_{\varphi,o}= 10\kmps$\,.
\end{itemize}
These values are typical of CTTS and close to the ones of RY Tau that we will study in more detail, except for the rotation rate,
which is much lower than the usual value. We argued, however, that the rotation rate of RY Tau is subject to caution
because of the obscuration of the star by the disk, as discussed below. Instead, we prefer to keep a value close to
that of most CTTS and remain more general.

Taking these constraints into consideration, we explored the parameter space and constructed various solutions  using the
procedure explained in STT02. Among the solutions adapted to T Tauri jets, we selected  two different sets of model
parameters that fit the RY Tau case particularly well. The first solution has a longer
lever arm, corresponding to our initial guess, and a large opening angle for the jet. This should be close to the observed values
for CTTS with a jet, such as the jet of RY Tau observed in February 2005 by \cite{StongeBastien08}. In this solution, the jet is
magnetically collimated into a cylindrical shape without any oscillation close to the star.

By a slight change of the pressure, we were able to get a second solution that is completely different from the
previous one. Of course, a change in one of the parameters implies a search for a new set of the other parameters to keep the
same observational constraints.
In this second solution, the lever arm is reduced, as is the opening angle. The solution exhibits strong recollimation and
oscillations. It may either reproduce WTTS or CTTS without jet such as RY Tau in 2001 when the jet was not visible. Instead, a UV 
shock was detected close to the star, possibly caused by the recollimation \citep{GomezVerdugo01}.

\section{Nonoscillating collimated solution}
\label{sec.4}

\subsection{Procedure}

In STT99 and STT02, we have shown that exact MHD solutions corresponding to efficient magnetic rotators are usually
collimated both by the magnetic field and by the external gas pressure. However, by decreasing the base pressure and keeping the
other parameters fixed,
we obtain a limiting solution where the asymptotic structure is collimated only by the magnetic hoop stress. From underpressured
close to the star, the jet becomes overpressured at large distances. For even lower base pressures the wind is no longer
collimated. We choose in this section a limiting solution that gives the widest possible opening of the jet. This can be 
done because the solution is an efficient magnetic rotator with a value of $\epsilon$ positive but close to zero (cf. STT99).
In addition to the parameters defined in Sect. \ref{sec.3.3}, to build our solutions we assumed that the jet speed at
infinity along the polar axis is $V_\infty = 200\kmps$.

The process of obtaining a self-consistent solution is a long iterating one. As explained in STT02, we start our integration at the
Alfv\'{e}n point and integrate upstream and downstream, each time crossing the critical points and changing the
parameters until we get as close as possible to the guessed values of the parameters.

\hspace{-0.5cm}\begin{figure}
\includegraphics[width= 9 cm] {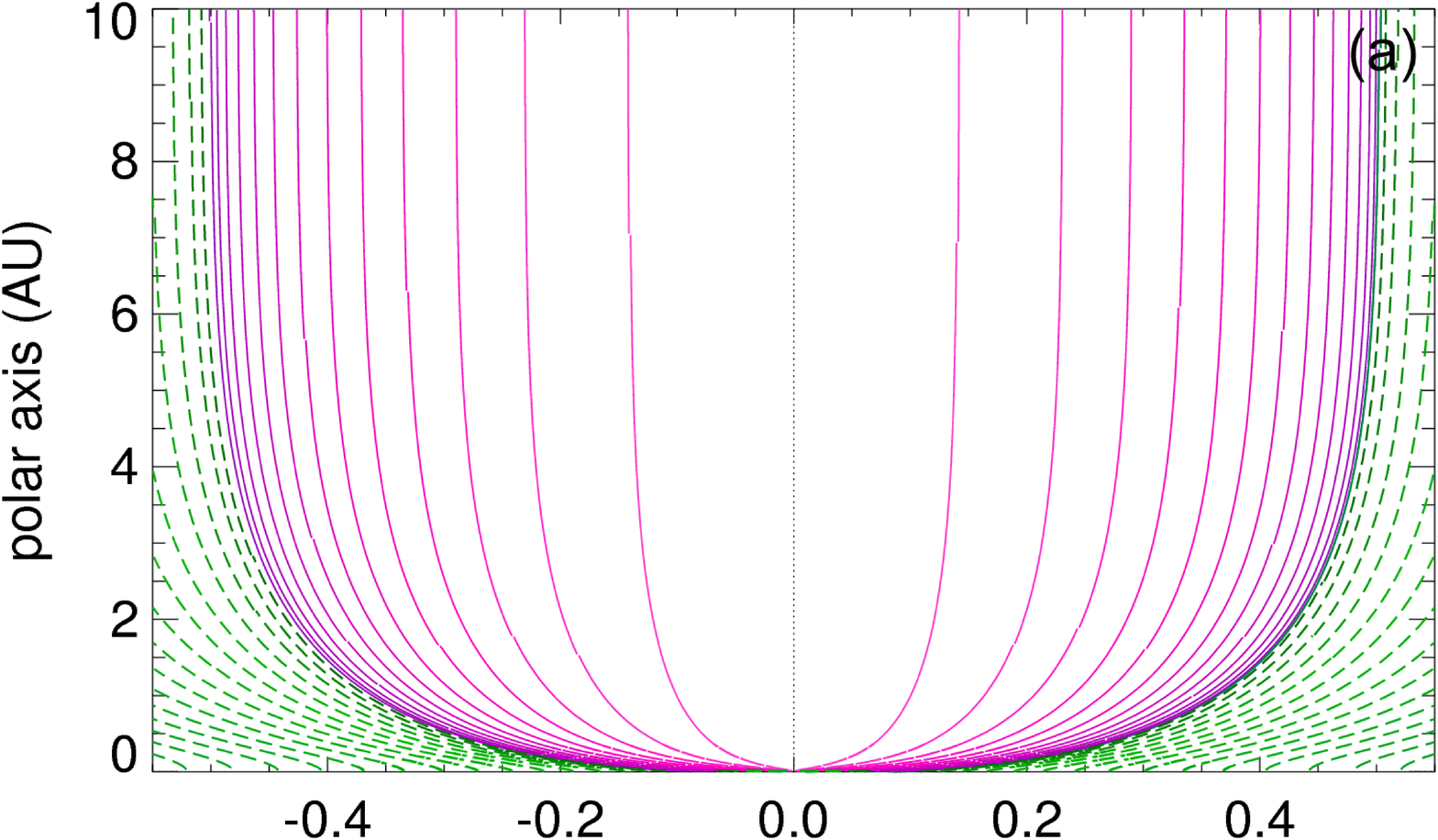}
\includegraphics[width= 9 cm] {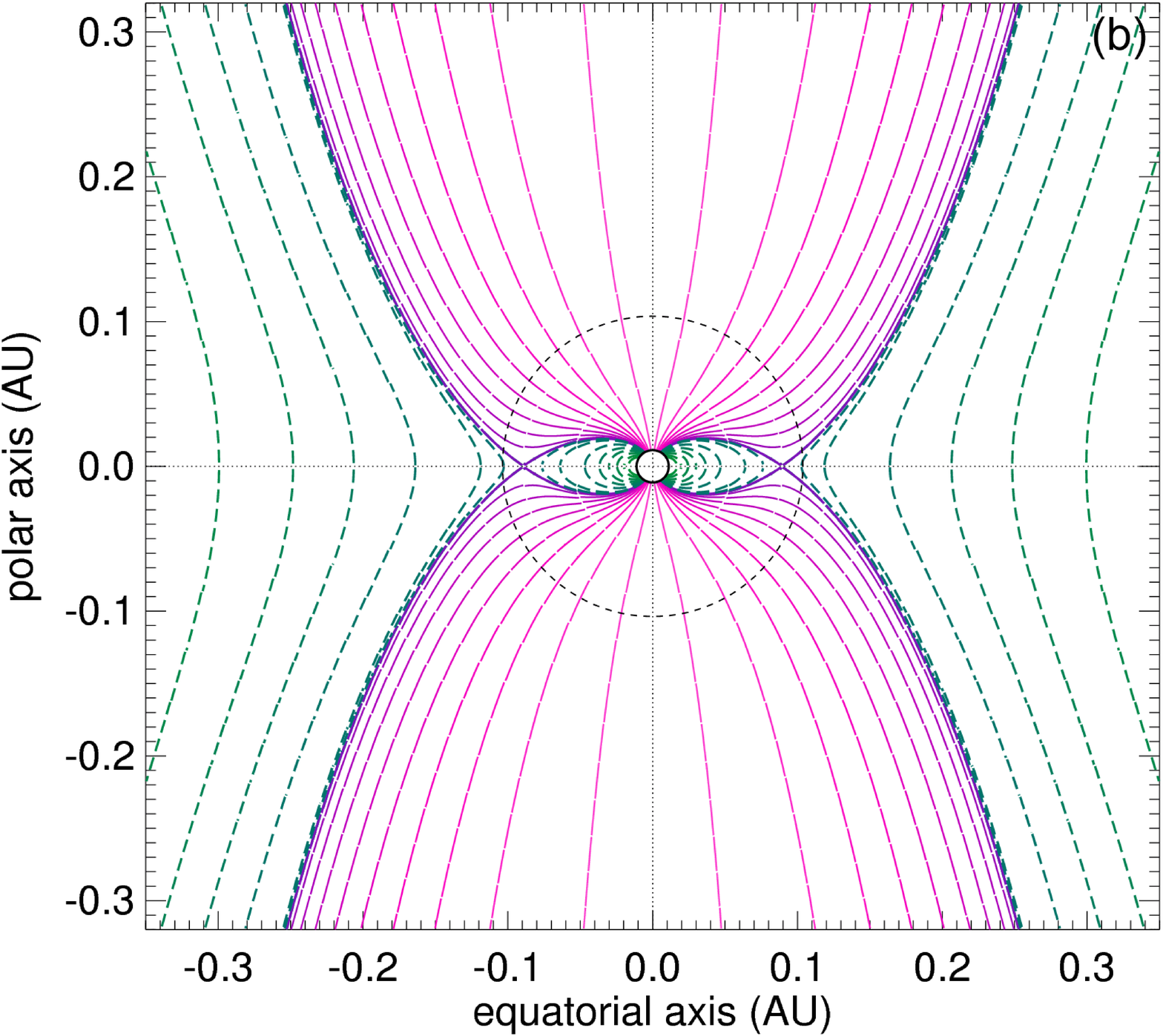}
\caption{
Topolopy of the nonoscillating solution in the meridional plane. In a) zoom out showing the collimation.
In b) zoom close to the star. All distances are given in astronomical units (AU). The solid (red) lines correspond to open fieldlines
connected to the star. The dashed (green) lines correspond to closed stellar fieldlines and lines connected to the disk. All open
lines are originating in a polar cap extending to  a colatitude of 15 degrees. The dotted circle indicates the Alfv\'{e}n surface.}
\label{Fig1}
\end{figure}

\subsection{A model for low mass accreting TTauri jets}

Our iterative procedure converges on the following set of parameters
\begin{itemize}
\item $\epsilon = 0.012$,
\item $\delta = 0.0778$,
\item $\kappa = 0.021$,
\item $\nu = 1.5$,
\item $\lambda = 0.775$.
\end{itemize}
with a rotation frequency of
\begin{itemize}
\item $\Omega = 5.15\,\, 10^{-6}$ rad/s, or, $V_{\varphi,o}= 8.6\kmps$,
\end{itemize}
This last value is in the lower range of the values measured for RY Tau. It corresponds to the value of a slowly rotating
T Tauri star. From this point of view the example of RY Tau is an extreme one.

There is a large discrepancy for RY Tau between determinations of the
spectroscopic projected rotational velocity and of the photometric rotational period. In this
particular case a value of $52$--$55\kmps$ is inferred for $V\sin i$ (\citeauthor{Petrovetal99}, \citeyear{Petrovetal99};
\citeauthor{Moraetal01}, \citeyear{Moraetal01}; \citeauthor{AgraAmboageetal09}, \citeyear{AgraAmboageetal09}),
where $i$ is the angle between the line of sight and the stellar rotational axis.   This value of $V\sin i$ is among  the highest
measured  for CTTS. This specific determination of the projected rotational velocity seems 
to correspond to a  special case of obscuration by the disk rather than a real rotational velocity, however. Thus, this value may
not correspond to the rotation of the star itself. Bouvier et al. (\citeyear{Bouvieretal93}, \citeyear{Bouvieretal95}) give a
photometric period of 24 days corresponding to $V_{\varphi,o}= 5.1\kmps$, which is a lower limit.
Flux measurements give a period from $5$ to $66$ days, closer to typical values for TTS
(\citeauthor{Petrovetal99}, \citeyear{Petrovetal99}). However, the variation is far from periodic and again
one cannot exclude veiling from the disk. In other words, rotation measurements are extremely difficult to obtain for this
object. We have tried to construct solutions with rotational values close to the canonical value of $15\kmps$, though within a
factor of $2$ or so, assuming that more precise observational data are needed.

\begin{figure}
\includegraphics[height=5.4cm]{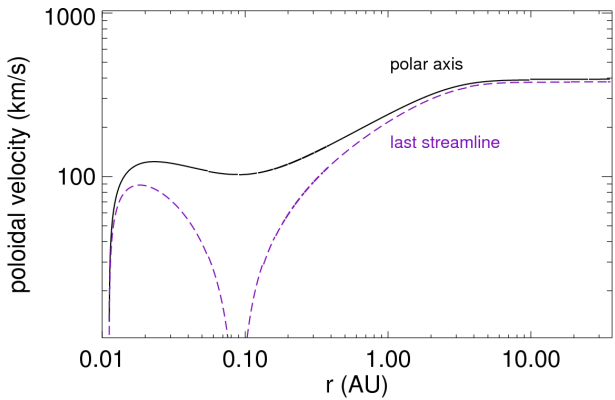}
\caption{
Plot of the poloidal velocity for the nonoscillating solution
along the polar axis (solid line) and the last streamline connected to the star (dashed). By construction the poloidal
velocity is zero on the last streamline where it reaches the equatorial plane. This creates the dip seen on the logarithmical scale. }
\label{Fig2}
\end{figure}

The topology of the streamlines and magnetic fieldlines of the solution in the poloidal plane is displayed in Fig. \ref{Fig1}.  The
stellar jet is surrounded by a diskwind and
is self-collimated by its own magnetic field. It is produced in a fairly narrow coronal hole of half opening angle 15 degrees. The
hole is surrounded by a large dead zone extending up to 8 stellar radii. This is comparable to the dipolar structure of 1.2 kG around
BP Tau reconstructed from ESPaDOnS observations \citep{Donatietal08}.
We also plot in Fig.  \ref{Fig2} the outflow speed along the polar axis and the one along the last open streamline connected to the
star. The asymptotic speed along the polar axis is
\begin{itemize}
\item $V_\infty = 393\kmps$\,,
\end{itemize}
which is higher by a factor of 2 than our initial guess. This may be quite large, but considering
the uncertainties involved, it can be an acceptable value.

\begin{figure}
\includegraphics[height=15.5cm]{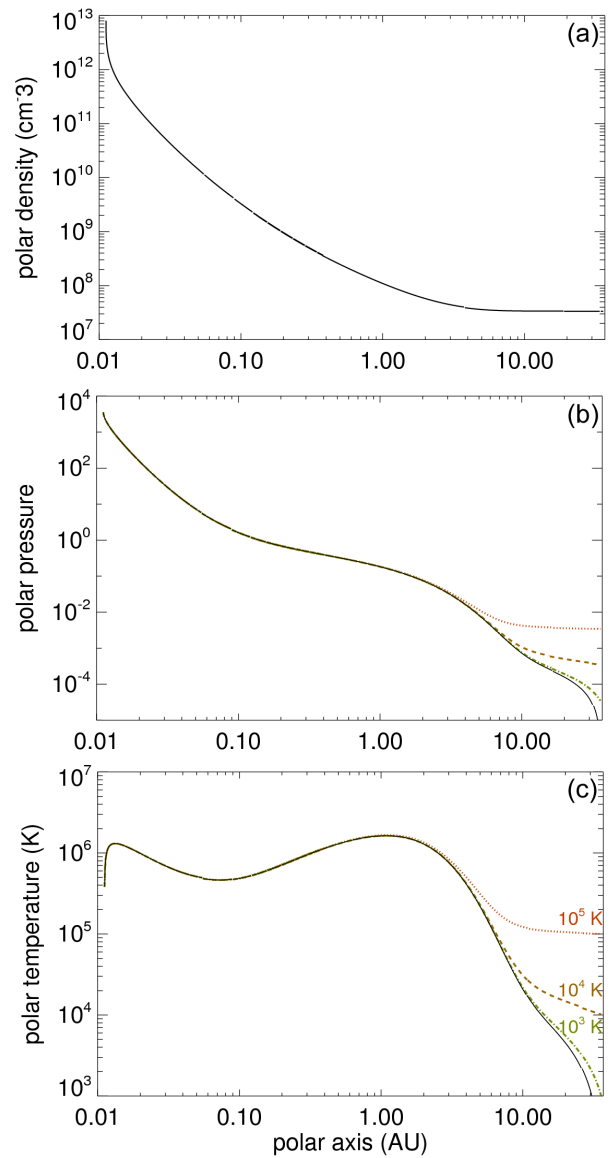}
\caption{
In (a) we plot of the polar density, in (b) the total pressure in dimensionless units along the polar axis, including nonkinetic terms  
such as
turbulent magnetic or ram pressure. In (c) we plot the polar effective temperature for the nonoscillating solution. The three additional
asymptotic curves in (c) correspond to three possible values of the temperature at infinity, obtained by adding non-zero values of the 
pressure at infinity, as shown in (b).}
\label{Fig3}
\end{figure}

In Fig. \ref{Fig3} we plot the polar density (a), pressure (b) and temperature (c). The pressure is defined up to a constant $P_o$, see
Eq. (\ref{pressure}). By varying this constant, we obtain the various curves of  Figs. \ref{Fig3}b and c. Adding different values of $P_o$
only affects the asymptotic part of the curves.

Note that the temperature in Fig. \ref{Fig3}c has a
maximum
along the polar axis, reaching a value of one million degrees, which might be a rather high value. However, this effective
temperature of the plasma is calculated
from the equation of state of the gas using the total pressure plotted in  Fig. \ref{Fig3}b. This total pressure may include in addition to
the kinetic pressure, ram pressure or Alfv\'en waves, etc. As we have shown in  \cite{Aibeoetal07}, the effective temperature can be
easily
ten times higher than the kinetic one with a relative amplitude of the Alfv\'en waves $\delta B /B$ less than unity. Note that
\cite{GomezVerdugo07} inferred from UV lines high electronic temperatures associated with a wind close to $10^5$K consistent
with this scenario and an effective temperature of one million degrees.

\hspace{-0.5cm}

\hspace{-0.5cm}


For this solution,  the physical quantities at the Alfv\'en radius are
\begin{itemize}
\item $r_\star = 9.29\,r_o = 0.104$ AU,
\item $V_\star = 103\kmps$.
\end{itemize}
The last open streamline connected to the star corresponds to a value of the dimensionless magnetic flux of
\begin{itemize}
\item $\alpha_{out} = 0.989$.
\end{itemize}
We assumed that the observed mass loss rate originates completely from the star. It means that the entire mass loss
rate is contained within a flux tube with its last line connected to the star.
This assumption yields a mass density, particle density and magnetic field at the Alfv\'en radius of
\begin{itemize}
\item $n_\star = 3.05\times 10^9\cmmt$
\item $\rho_\star = 2.48\times 10^{-15}\gcmmt$
\item $B_\star = 1.82$ G
\,,
\end{itemize}
respectively, where $\rho_\star$  is given by (see Eqs. (3.14) and (3.15) in ST94)
\begin{equation}
\rho_\star
=\frac{\dot M_{\rm wind}}{2\pi r_\star^2 V_\star \psi_{\rm out}}
= \frac{3\delta \dot M_{\rm wind}}{4\pi r_\star^2 V_\star}\frac{1}{(1+\delta\alpha_{\rm out})^{3/2}-1}
\label{rhostar}
\,.
\end{equation}
We recall that the mass loss rate of this solution is $\dot M_{\rm wind} = 3.1\times 10^{-9}M_\odot/$yr.
From these values at the Alfv\'en  radius we obtain
\begin{itemize}
\item $\rho_\infty = 2.72\times 10^{-17}$g\,cm$^{-3}$, deduced from $\rho_\star$ and the Alfv\'en Mach number at infinity,
\item $n_\infty \approx 10^7$cm$^{-3}$ assuming the plasma is fully ionized,
\item $B_o = 608$ G,
\item $B_\infty = 76$ mG,
\item $\varpi_{out}(\frac{\pi}{2}) = 7.97 \, r_o =0.0891$ AU,
which corresponds to the  cylindrical radius where the last line connected to the star reaches the equatorial plane
(see Fig. \ref{Fig1} b)
\item $\varpi_{\infty, out} = 42.8 \, r_o = 0.478$ AU,
\end{itemize}
which corresponds to the radius of the jet at large distances. The radii of bow shocks in jet simulations are much larger than the jet itself. They
can easily be 10 times larger (e.g. \citeauthor{Matsakosetal09}, \citeyear{Matsakosetal09}). A jet with a radius of 1 AU may create
shocks  of width 10 AU. Thus
the radius we have obtained in this solution is consistent with current observations, though for low mass accreting stars the jet
radius cannot be resolved \citep{StongeBastien08}.

The star-braking time calculated from Eq. (\ref{tau})  for this solution is
\begin{equation}
\tau  \, \approx\,0.6 \times10^{6} yr
\label{tauCTTS}
\,.
\end{equation}
This corresponds to the typical life time of a CTTS \citep{Bouvieretal97}.
If $k$ equals 0.2 instead of 0.06, we obtain a star braking time of 2 million years.
This does not drastically affect our final conclusion. This means that the jet is efficiently
removing angular momentum from the star. Thus we may explain the low rotation frequency of those particular stars without
invoking the disk locking mechanism. This calculation relies on the steady solution we used. This does not prevent the
star from sudden violent outflows caused by reconnection as shown in various numerical simulations. As we mentioned
in the introduction, the presence of coronal mass ejections (CMEs) does not affect the longterm evolution of the steady flow.
It is interesting, however, to note that the braking time obtained in the \cite{Kuekeretal03} simulations is very similar, albeit with a
different process.  The authors conclude that the accretion torque spins the star up because the  magnetic torque from the X-wind is too
weak. We propose here alternatively that the stellar jet could compensate for it.

\subsection{Higher mass accreting TTauri jets}

In brighter T Tauri jets, higher mass loss rates are usually observed. We may extend the same solution in cylindrical
radius by including the disk-wind component. To obtain a jet  mass
loss rate of $\dot M_{\rm wind} = 10^{-8}M_\odot/$yr,  the solution should be extended to
include the fieldline region connected to the disk up to $\alpha_{disk, out}  = 3.07$.
Note that this value of $\dot M_{\rm wind}$  is in the range of the mass loss rates, deduced  for
RY Tau seen in the optical by \cite{AgraAmboageetal09}, typically  $0.16$ to $2.6 \times 10^{-8}M_\odot/$yr.

The extended jet has a launching region of radius
$\varpi_{disk, out} = 0.478$ AU and an asymptotic cylindrical radius of
$\varpi_{\infty, out} = 0.890$ AU. The extension of the launching region within the disk is quite large. Conversely the expansion of
the disk outflow is too small.  Indeed, we do not expect our model to be
extended to  this distance from the
star.  It does not apply to the Keplerian part of the accretion disk region. A more realistic disk wind model would naturally give a
larger expansion factor of the jet and an asymptotic radius of a few tens of AU, as expected observationally.

We can also keep the same solution considering only  the stellar
wind component  but assuming a mass loss rate of
$\dot M_{\rm wind} = 10^{-8}M_\odot/$yr.  Accordingly we recalculate the physical quantities.
Increasing the mass loss rate by a factor of 3 just increases the density by a factor of 3 and the magnetic field
by a factor of $\sqrt{3}$. This does affect the velocity and the temperature within the jet. In this case the values at the Alfv\'en
radius are
\begin{itemize}
\item $n_\infty \approx 3.2 10^7$cm$^{-3}$ assuming the plasma is fully ionized,
\item $B_o \approx 1.1$ kG,
\item $B_\infty  \approx 140$ mG.
\end{itemize}
This adapts our solution to higher mass loss rates. In these cases however,  the large observed jet radii imply that there is a 
significant  disk wind contribution. Note that this high mass loss rate changes the braking time and reduces it by a factor of 3
\begin{equation}
\tau \,\approx \,2 \times10^{5} yr
\label{tauCTTS}
\,.
\end{equation}
It also increases the dipolar magnetic field at the surface of the star up to 1.1 kG, which is still reasonable \citep{Donatietal08}. It does
not change the temperature profile of Fig. \ref{Fig3}, but because the density is significantly higher, this would probably lead to
observational signatures in X-rays.

\section{A recollimating solution}
\label{sec.5}

\subsection{Construction of the solution}
\label{sec.5.1}

\begin{figure}
\includegraphics[width=9cm]{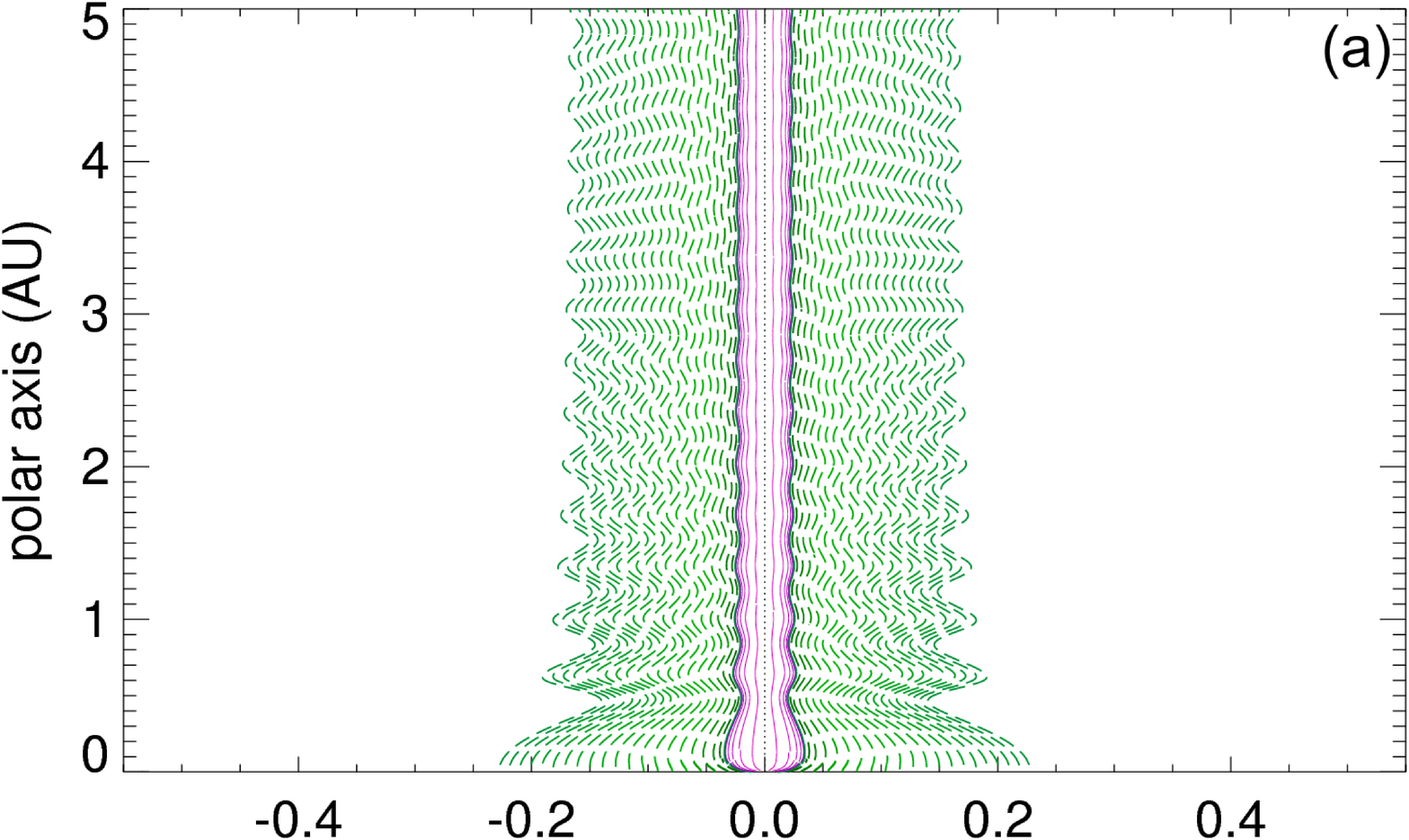}
\includegraphics[width=9cm]{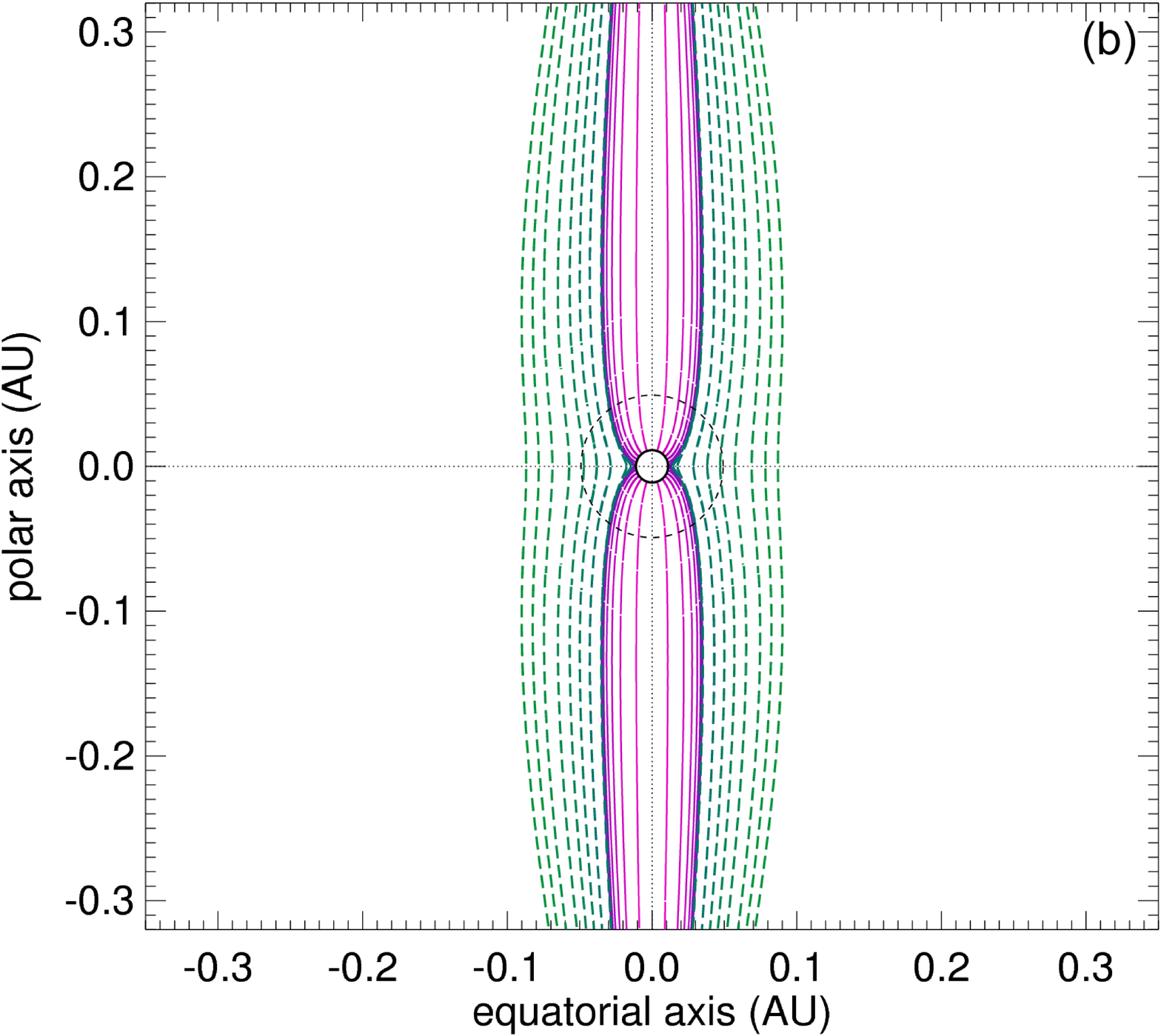}
\caption{Topolopy of the streamlines and magnetic fieldlines of the oscillating solution in the meridional plane. In a) zoom out
showing the recollimation.
In b) zoom close to the star. All distances are given in astronomical units (AU). The solid (red) lines correspond to open fieldlines
connected to the star. The dashed (green) lines correspond to lines connected to the disk. The
dead zone here is reduced and all lines from the star are opened and not connected to the accretion disk. The solution
recollimates at 38 stellar radii at a distance where the UV shock was observed for RY Tau from 1993 to 2001. The dotted circle indicates the Alfv\'{e}n surface.}
\label{Fig4}
\end{figure}

By changing the initial pressure and the other parameters accordingly to keep the same stellar constraints, we obtain a second
solution, which recollimates at about 38 stellar radii.  The morphology of
this solution is displayed in Fig. \ref{Fig4}.

The model
parameters are close to those corresponding to RY Tau again and the recollimation distance in that case corresponds to the
distance of the shock seen in the UV by \cite{FerroFontanDeCastro03}.
From the HST observations obtained in 1993, \cite{GomezVerdugo01} deduced the presence of a UV shock and suggested that it
might correspond to the recollimation of the underlying jet.
Following this track, we suggest that a shock could form in our solution at the position where the jet recollimates. Of course, a
simple analytical model cannot describe the shock itself, this is beyond the scope of the present paper. We just note here that the
outflow velocity at this point is super fast magnetosonic, and hence shocks can form without destroying the outflow. This has been shown numerically
by \cite{Matsakosetal09} for this solution. 

\cite{StongeBastien08} observed episodic small scale outflows in 1998 and 2005 always along the same ejection axis. They
proposed that these successive knots merge
to form the large-scale jet. We suggest then that the two solutions may correspond to two different
stages of the same object. The recollimating solution may correspond to the stage with a UV shock and the nonoscillating 
solution to another phase where the jet is visible.

\hspace{-0.5cm}

\subsection{Physical discussion on the oscillating solution}

After iterating, we found the following set of parameters
\begin{itemize}
\item $\epsilon = -0.034$\,,
\item $\delta = 0.075$\,,
\item $\kappa = 0.065$\,,
\item $\nu = 5.8$\,,
\item $\lambda = 0.884$\,.
\end{itemize}
Note that the value of $\epsilon$ is now slightly negative. This means that the magnetic rotator is important but less efficient in 
collimating. We know from previous study (STT99) that in that case the oscillations are more pronounced.

\begin{figure}
\includegraphics[height=5.4cm]{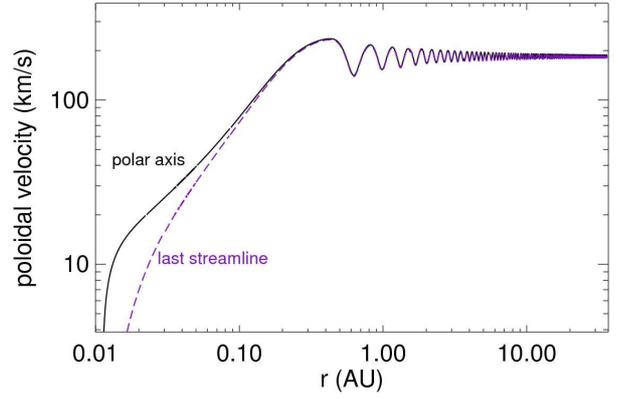}
\caption{
Plot of the poloidal velocity for the oscillating solution
along the polar axis (solid line) and the last streamline connected to the star (dashed). The jet decelerates after the
recollimation distance. }
\label{Fig5}
\end{figure}

\begin{figure}
\includegraphics[height=15.5cm]{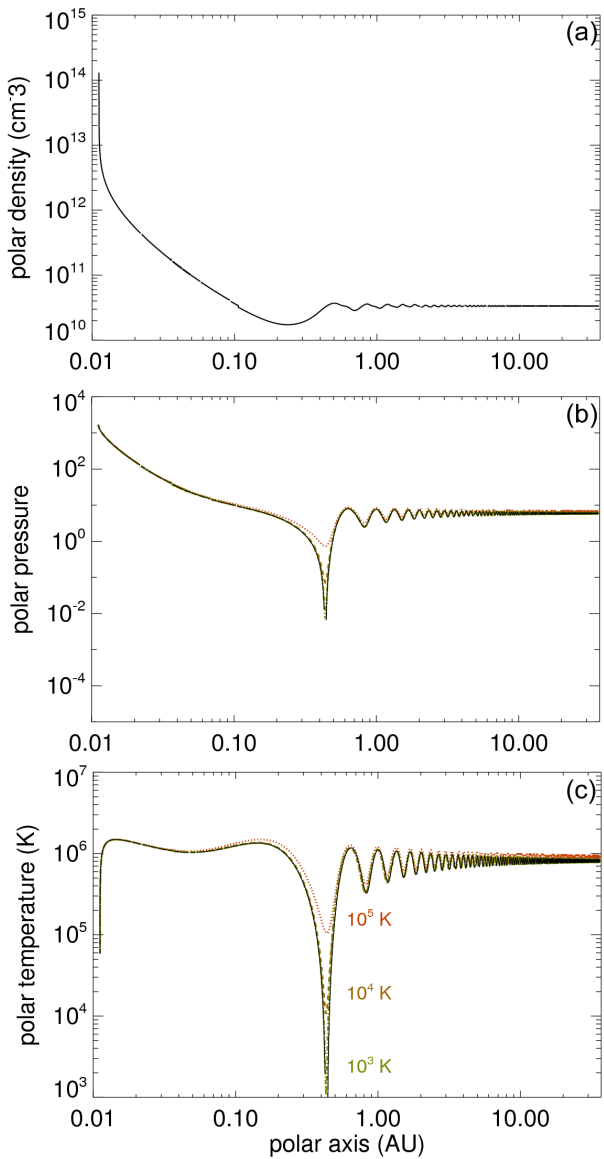}
\caption{
Plot of the polar density (a), the dimensionless polar pressure (b) and the polar temperature (c) for the oscillating solution.
The temperature is the effective one as in Fig. \ref{Fig3}. The polar effective temperature is fairly constant along the
jet, suggesting a uniform level of turbulence. The three extra asymptotic curves in (c) correspond to three possible values of the 
temperature at infinity, obtained by subtracting non-zero values of the pressure at infinity, as shown in (b).}
\label{Fig6}
\end{figure}

The corresponding rotation frequency, azimuthal velocity and asymptotic speed along the polar axis
are
\begin{itemize}
\item $\Omega = 4.68\,\times\, 10^{-6}$ rad/s\,,
\item $V_{\varphi,o}= 7.82\kmps$\,,
\item $V_\infty = 186\kmps$\,,
\end{itemize}
 respectively. 
 This last value is closer this time to our initial guess for RY Tau. We plot the polar velocity of this solution in Fig.  \ref{Fig5},
the corresponding polar density, pressure and temperature in  Fig.  \ref{Fig6}. Similarly to Figs. \ref{Fig3}b and c, changing the  value
of $P_o$ only affects the curves near their minimum but not close to the star.

\hspace{-0.5cm}

\hspace{-0.5cm}

For this solution,  physical quantities at the Alfv\'en radius are:
\begin{itemize}
\item $r_\star = 4.39\, r_o = 0.049$ AU
\item $V_\star = 38.75\kmps$
\,.
\end{itemize}
We calculate the density again assuming that the observed mass loss rate of $\dot M_{\rm wind} = 3.1\times 10^{-9}M_\odot/$yr
entirely comes from the star. This means that the jet radius is defined as the last flux tube connected to the star that
corresponds to a magnetic dimensionless flux
\begin{itemize}
\item $\alpha(f_{min})=\alpha_{out} = 0.361$ (see Fig. \ref{Fig4}).
\end{itemize}
It corresponds to a mass density, particle density and magnetic field at the Alfv\'en radius of, respectively,
\begin{itemize}
\item $n_\star = 1.01\times 10^{11}\cmmt$ is we assume a fully ionized plasma as previously\,,
\item $\rho_\star =8.19\times 10^{-14}\gcmmt$\,,
\item $B_\star = 3.94 $ G\,.
\end{itemize}

From this solution we obtain the following output values
\begin{itemize}
\item $B_o = 27.4$ G\,,
\item $B_\infty = 6.28$ G\,,
\item $\varpi_{\infty; out} = 2.09\,r_o = 0.023$ AU\,.
\end{itemize}
The asymptotic density is higher in this solution because we kept the same mass loss rate, but both the velocity and the jet radius are
smaller. However, the stellar jet radius is so small in this case that even if it were drastically enlarged by a shock, say at most
by a factor of ten, it would remain far below any possible detection.

This solution, as we mentioned, may describe RY Tau or any other CTTS during phases where the jet is not visible. We suggest
that it may also be well adapted to WTTS where there is no obvious jet observed despite the similarity of the central star.

Moreover, the star braking time calculated from Eq. (\ref{tau}) is for this solution
\begin{equation}
\tau  \, \approx\,7.7\times10^{6} yr
\label{tauWTTS}
\,.
\end{equation}
This is longer than the lifetime of CTTS, closer to the time spent by the star once the disk is dissipated. Thus the magnetic
braking is strongly reduced in this case and inefficient. If the accretion rate remains identical, one would expect  a spin-up
of the star. But it is unclear in this solution if the accretion rate is not also reduced because of the reduction of the
large dead zone of the first displayed solution (compare Fig. \ref{Fig1}b and \ref{Fig4}b).

\subsection{A possible application to the RY Tau jet}

By analyzing high spatial resolution images of the classical T Tauri star RY Tauri in the Taurus-Auriga cloud taken with
the Gemini North telescope, \cite{StongeBastien08} strongly suggested that the RY Tau jet is a pure stellar jet. They argue that the
knots are  intermittent microjets that merge on longer scales. In 2005, they deduced the tangential velocity of the brightest knot
already seen by HST in 1998. Parallely, in 1993 and 2001, a UV shock was observed
by \citeauthor{GomezVerdugo01} (\citeyear{GomezVerdugo01}, \citeyear{GomezVerdugo07}). They interpreted it as a signature of a
stellar jet component.

In view of these observations, it is tempting to suggest that the two different MHD solutions discussed above may
correspond to the two different stages of RY Tau.
Thus, to explain the succession of knots seen close to the star, the first nonoscillating solution would be relevant for this stage.
During this period,
the stellar activity of the central star could be enhanced, producing a large visible stellar jet that efficiently brakes the central star.

On the other hand, during periods where the UV lines are detected, the outflow could be described by the second oscillating solution
with an invisible jet, except for the recollimating region where the UV shock is observed. Moreover, the width of the stellar jet at the
recollimating point is around 2 stellar radii, which corresponds to the width of the UV zone
before the shock, as observed. This may take place during a short period, probably linked to a lack of magnetic activity, or even
during stellar minimum.

The jet alternately takes the morphology of the two solutions depending on the activity of the star. A UV shock is formed during the
low activity phase. Once the activity is high, the jet speed increases and blows away the  UV shock.

Note that this multiphase behavior of the large scale jet has been found in time-dependent MHD simulations of
a two-component MHD outflow, when the initial injection speed of the jet varies \citep{Matsakosetal09}, a variability that may be 
caused by a magnetic activity cycle.

The density profiles of our two solutions are consistent with the densities given by \cite{GomezVerdugo07}.  They measure an
electronic  density of $10^{12}$ cm$^{-3}$ in the SiIII UV line between 0.006 and 0.3 AU, and $10^{10}$ cm$^{-3}$ in the CIII] UV
semi-forbidden line between 0.05 and 1 AU. These two values are consistent with the densities of the second solution displayed in
Fig. \ref{Fig6}a.
They also measure a density of  $10^7$ cm$^{-3}$ in the [OII] forbidden UV line farther out between 4 and 100 AU. This last line is
not formed in the UV shock region. The observed density corresponds to the asymptotic density of the first solution displayed in
Fig. \ref{Fig3}a.

\section{Conclusions}
\label{sec.6}
We have used a simple semi-analytical MHD model (ST94) to show that the low rotational velocity of T Tauri stars
can be understood by considering the effect of stellar jets to remove their angular momentum.
In particular, we modeled stellar jets from CTTS with a mass loss rate consistent with observations. The solutions
for collimated magnetized outflows are obtained via a nonlinear separation of the variables in the governing full set of the MHD
equations.
Two main results are obtained.


{\it First}, jets can efficiently brake the central star and remove the bulk of its angular momentum over a timescale of a million
years, at least for low mass accreting stars. This corresponds to the typical lifetime of  a star in the CTTS phase.
Furthermore, with this mechanism the star can even slow down within 0.6 million years, if the disk does not transport angular
momentum onto the star.
This fairly short slow-down time shows the efficiency with which a magnetized outflow can remove angular momentum from a
rapidly rotating young stellar object.
This result strongly suggests that stellar jets may explain the low rotational speeds observed in those objects, independently of
any disk locking. This mechanism could be complementary to other angular momentum removal mechanisms occurring at the X- 
wind emerging from the
interaction zone between the disk and the magnetosphere.
Our solution fairly well reproduces the mass loss rate, terminal speed and rotation rate of the jet. The effective plasma
temperature in the jet is reasonable provided that non thermal processes, such as Alfv\'en waves, are at work in the plasma of
the jets. These
physical processes are likely to take place in the jets if one considers that the magnetic activity of CTTS is comparable to, or higher
than, that of our own Sun.


{\it Second}, we also analyzed two interesting MHD solutions. The first corresponds to a  gradually cylindrically
collimated wide wind without oscillations in its width. The second is a
narrower outflow that refocuses toward the jet axis and also oscillates in width. 
Note that these two very different solutions were obtained by a slight change of the parameters. 
It is a known result in nonlinear equations (e.g. Parker solutions, or, their MHD generalization).
Correspondingly, we discussed the applicability of these two solutions in the context of the RY Tau jet, of which
recent observations indicate that the plasma outflow from the rotating and magnetized star may have an intermittency exhibiting
at least two phases. During one phase the cylindrical wind is relatively wide and nonoscillating.
During the other phase the outflow is narrow and oscillates.
Oscillations may produce radiative shocks seen in various emission lines, such as the observed UV lines of RY Tau.
\cite{Matsakosetal09} have shown that the self-similar solutions used in this paper will not be destroyed by the shock formation or
the disk-wind interaction.

From a more general perspective, this dichotomic behavior can be seen in the two classes, CTTS and WTTS, which may
both actually have jets, but these jets are visible only CTTS. The connection of CTTS with their disks through their magnetospheres
may increase the width of their jets such as to make them visible.
The mass loss rate measured in CTTS is thought to decrease as the star evolves toward the stage of WTTS. In this later stage,
direct
observations are more difficult and upper limits around $10^{-10}M_\odot/$yr can only be inferred. Weak T Tauri stars seem to have lost the gas component of their accretion disk or at least gas falling onto the star. Our second MHD
solution suggests that even with the same mass loss rate, the jets of WTTS may be invisible with the present angular resolution
because of their narrowness. The lack of accretion disk prevents the formation of disk winds and the possibility of higher mass loss rates.
The lower jet width may be caused either by a lower stellar magnetic activity (see \citeauthor{Vidottoetal09}, \citeyear{Vidottoetal09},
\citeyear{Vidottoetal10}) or by
a lack of magnetic connection into the disk. The magnetic braking time in this case would be on the order of 10 million years,
which is indeed the time a T Tauri star is expected to spend in the weak-line stage.

Our modeling suggests that CTTS may produce visible jets in their active phase that would efficiently brake the star. Indeed, in
the nonoscillating
solution the magnetic field dominates the dynamics in the jet. Near the star, the magnetic field has a dipolar structure and the
wind is more efficient in this case in extracting stellar angular momentum.
On the other hand, WTTS may have jets as well, which are invisible because they are too narrow.
These jets would be weaker because of the lack of direct connection with the accretion disk through a large dead zone.


\begin{acknowledgements}
The authors are grateful to J.F. Gameiro for his useful comments during the preparation of the manuscript and
acknowledge support through the Marie Curie Research Training Network JETSET (Jet Simulations, Experiments and Theory)
under contract MRTN-CT-2004-005592. ZM acknowledges financial support from the FWO, grant G.0277.08. JJGL acknowledges
financial support from project PTDC/CTE-AST/65971/2006 from FCT, Portugal and the financial support and kind hospitality
of the Observatory of Paris. VC, NG and CS thank CAUP and the University of Athens for their kind hospitality.

\end{acknowledgements}

\appendix

\section{Governing equations for meridional self-similar outflows}
\label{A}

The basic equations governing plasma outflows in the framework of an ideal MHD
treatment for steady, axisymmetric flows are the  momentum, mass and magnetic
flux conservation, the frozen-in law for infinite conductivity,
and the first law of thermodynamics. In particular, the poloidal component
of the magnetic field can be derived from the magnetic flux in spherical
coordinates ($r, \theta, \varphi$),
\begin{equation}\label{B}
\vec{B}= {\vec \nabla A \over r \, \sin \theta } \times \hat \varphi
\,.
\end{equation}

We need to specify the latidudinal dependences of the velocity, magnetic,
density and pressure fields, $\vec{V}$, $\vec{B}$, $P$ and $\rho$ respectively.
This can be resumed in the following assumptions (for details see
ST94, STT99 and STT02):
\begin{equation}
\label{Br}
B_r =B_{*} {1\over G^2(R)}\cos\theta\,,
\end{equation}
\begin{equation}
B_\theta =-B_{*} {1\over G^2(R)}{F(R)\over 2}\sin\theta
\,,
\end{equation}
\begin{equation}
\label{Bphi}
B_\varphi = - B_{*} {\lambda \over G^2(R)}
{\displaystyle 1 - G^2(R) \over 1 - M^2(R) }{R\sin\theta}
\,,
\end{equation}
\begin{equation}
\label{Vr}
V_r = V_{*}  {M^2(R)\over G^2(R)} { \cos\theta \over
\sqrt{1+\delta \alpha(R,\theta)}  },\;\;\;\;\;\;
\end{equation}
\begin{equation}
V_\theta =-V_{*} {M^2(R)\over G^2(R)}{F(R)\over 2}
{  \sin\theta \over \sqrt{1+\delta \alpha(R,\theta)}  }
\,,
\end{equation}
\begin{equation}
\label{Vphi}
V_\varphi = V_{*}  {\lambda \over G^2(R)}
{ G^2(R) - M^2(R) \over 1- M^2(R)}
{R\sin\theta \over \sqrt{1+ \delta \alpha(R,\theta) } }
\,.
\end{equation}
\begin{equation}\label{density}
\label{rho}
\rho(R,\alpha) = {{\rho_*} \over {M^2(R)}} (1 + \delta \alpha)
\,.
\end{equation}
\begin{equation}\label{pressure}
P(R,\alpha) = {1 \over 2} \rho_* V^2_* \Pi(R)[1+ \kappa \alpha]
+P_o
\,.
\end{equation}

We used
\begin{equation}\label{M}
M^2 \equiv M^2(r)= 4 \pi \rho {{V^2_p} \over {B^2_p}}
\,,
\end{equation}
the square of the poloidal Alfv\'en number, which is a function of
radial distance only.

We defined the dimensionless magnetic flux function
$\alpha(R, \theta) = 2 \, A(r, \theta) / r^2_* B_*$.
This quantity is related to $G(R)$ through the following expression
\begin{equation}\label{alpha}
\alpha = {{R^2} \over G^2(R) } {\rm sin}^2 \theta
\,,
\end{equation}
\noindent
where $G^2 (R)$ is the cross-sectional area of a flux tube perpendicular to
the symmetry axis in units of the corresponding area at the Alfv\'en distance.

Finally, for homogeneity with the notations in  ST94, STT99 and STT02,
we also introduced the  function $F(R)$, which is the logarithmic
derivative (with a minus sign) of the well known expansion factor used in
solar wind theory (\citeauthor{KoppHolzer76}, \citeyear{KoppHolzer76}):
\begin{equation}\label{F}
F(R) = 2 \, \left [1 -
{{\mathrm d} \, \ln G(R) \over {\mathrm d} \, \ln R} \right ]
\,.
\end{equation}
We recall that the value of $F$ defines the shape of the
poloidal streamlines.
For $F(R) =  0$ the streamlines are radial, for $F(R) > 0$ they are
deflected to the polar axis ($F=2$ means cylindrical collimation),
while for $F(R) < 0$  they flare to the equatorial plane.

Using these assumptions, the usual MHD equation system reduces to the three following ordinary differential equations for $\Pi(R)$,
$M^2(R)$ and $F(R)$,
\begin{equation}\label{Eq1}
{\hbox {d} \Pi \over \hbox {d} R}=
- {2 \over G^4 }
  \left[ {\hbox{d} M^2 \over \hbox{d} R} + {M^2 \over R^2} (F-2) \right]
-  {\nu^2 \over M^2 R^2 }
\,,
\end{equation}
\begin{equation}\label{Eq2}
{\hbox {d} F(R) \over \hbox {d} R}={{{\cal N}_F(R,G,F,M^2,\Pi; \kappa, \delta,
\nu, \lambda)} \over \     {R \, {\cal D}(R,G,F,M^2; \kappa, \lambda)}}
\,,
\end{equation}
\begin{equation}\label{Eq3}
{\hbox {d} M^2(R) \over \hbox {d} R}={{{\cal N}_M(R,G,F,M^2,\Pi;
\kappa, \delta, \nu, \lambda)} \over \     {R \, {\cal D}(R,G,F,M^2; \kappa,
\lambda)}}
\,,
\end{equation}
where we defined
\begin{equation}\label{Eq4}
{\cal D}= (M^2-1)\left( 1+\kappa {R^2\over G^2} \right)
         + {F^2\over 4} + R^2\lambda^2{N_B^2\over D^2}
\,,
\end{equation}
\begin{eqnarray}\label{Eq5}
{\cal N}_F = -(\delta-\kappa)\nu^2 {R G^2\over 2 M^2}F
\nonumber\\
+ \left({2\kappa \Pi G^2 R^2} + (F+1)(F-2) \right)
\times
\nonumber\\
\times
\left(1+\kappa {R^2\over G^2} - {F^2\over 4}
                      -R^2\lambda^2{N_B^2\over D^3} \right)
\nonumber\\
+{M^2F\over4}(F-2)\left(F+2+2\kappa{R^2\over G^2}
            +2R^2\lambda^2{N_B^2\over D^3} \right)
\nonumber\\
-\lambda^2 R^2 F(F-2){N_B\over D^2}
\nonumber\\
+\lambda^2 R^2 \left(1+\kappa {R^2\over G^2} -R^2\lambda^2{N_B^2\over D^3}
                    - {F\over 2} \right)
\nonumber\\
\left( 4{N_B^2\over D^2}-{2\over  M^2}{N_V^2\over D^2}\right) \,,
\end{eqnarray}
\begin{eqnarray}\label{Eq6}
{\cal N}_M= (\delta-\kappa)\nu^2 {R G^2\over 2 M^2}(M^2-1)
\nonumber\\
+ \kappa \Pi R^2 G^2 M^2{F\over 2}
-{M^4\over 4}(F-2)(4\kappa {R^2\over G^2} +F+4)
\nonumber\\
+{M^2\over 8}(F-2)(8\kappa {R^2\over G^2} +F^2+4F+8)
\nonumber\\
- \lambda^2 R^2 (F-2){N_B\over D}
\nonumber\\
+\lambda^2 R^2 (2M^2+F-2)\left(
{N_B^2\over D^2} -{1\over 2 M^2}{N_V^2\over D^2}\right)
\,,
\end{eqnarray}
with
\begin{equation}\label{Eq7}
N_B = 1 - G^2,\;\;\;\;N_V=M^2-G^2,\;\;\;\;D=1-M^2
\,.
\end{equation}
The meaning of the various parameters is discussed in Sec. 2.

At the Alfv\'en radius, the slope
of $M^2(R=1)$ is $p= (2 - F_*)/ \tau$, where $\tau$ is a solution of the
third-degree polynomial:
\begin{equation}\label{Eq8}
\tau^3 + 2 \tau^2 + \left [ {{\kappa \Pi_*} \over {\lambda^2}} +
{{F^2_* - 4} \over {4 \lambda^2}} - 1 \right ] \tau +
{{(F_*-2)F_*} \over {2 \lambda^2}} = 0
\,,
\end{equation}
and  the star indicates values at $R=1$ (for details see ST94).


\end{document}